# Pattern Evolution Characterizes the Mechanism and Efficiency of CVD Graphene Growth

Wanzhen He[1], Dechao Geng[2] and Zhiping Xu[1,*]

[1]Applied Mechanics Laboratory, Department of Engineering Mechanics and Center for Nano and Micro Mechanics, Tsinghua University, Beijing 100084, China.

[2]Pillar of Engineering Product Development, Singapore University of Technology and Design, 8 Somapah Road, Singapore 487372, Singapore

[*]Corresponding author, email: xuzp@tsinghua.edu.cn

## Abstract

Growing large-area, high-quality monolayers is the holy grail of graphene synthesis. In this work, the efficiency of graphene growth and the quality of their continuous films are explored through the time evolution of individual domains and their surface coverage on the substrate. Our phase-field modeling results and experimental characterization clearly demonstrate the critical roles of the deposition flux, edge-reaction kinetics and the surface diffusion of active carbon sources in modulating the pattern evolution and rate of growth. We find that the contrast between the edge-kinetics-limited and surface-diffusion-limited regimes is remarkable, which can be characterized by the evolution of domain patterns and considered as an indicator of the growth regime. However, common features exist in these two regimes, showing that the growth rate scales with time as $t^2$ in the early stage of growth and is regime-independent, which is explained by the coarsen profiles of carbon concentration for both the compact and dendritic domains. The rate decays rapidly in the final stage of growth due to the competition between neighboring domains on the limited carbon sources diffusing on the substrate, which is highly regime-sensitive and extremely low in the surface-diffusion-limited regime with narrow gaps between the domains to be filled. Based on these findings, synthesis strategies to improve the growth efficiency and film quality are discussed.

Graphene, a representative two-dimensional (2D) material, has attracted significant attention in recent years because of its exceptional material properties, which can further be engineered in various ways.[1] The crystallinity of large-scale graphene monolayers is of critical importance for high-profile applications, in optoelectronics for example. Mechanical exfoliation and chemical vapor deposition (CVD) are the two common techniques to obtain high-quality graphene samples.[2-3] Unfortunately, graphene monolayers cleaved from graphite are usually polycrystalline with the grain size of tens to hundreds of micrometers.[4] On the other hand, the size of a single-crystalline domain in CVD-grown graphene is limited to a few centimeters before it merges with neighboring domains, where line defects containing topological defects could form at the interfaces resulted from the misalignment if the true epitaxy condition cannot be assured. It usually takes tens of minutes to complete the whole process of film completion. Meter-size single crystals were recently synthesized on single-crystal Cu (111) surface.[5-6] Graphene domains nucleated at multiple sites were highly aligned, resulting in seamless, defect-free interfaces between the domains in coalescence, and a continuous film of 5 cm x 50 cm could be obtained after a fast growth process of 20 minutes. However, there are evidences showing that the growth of graphene slows down and the time cost for the coalescence of a full graphene monolayer is significant.[7-9] Although intensive studies were conducted on the domain pattern evolution during the CVD growth,[10-13] concluding that continuous graphene monolayers were formed, this evolutional pathway of film completion from individual domains has rarely been discussed, and the efficiency of growth as well as the quality of as-grown film thus cannot be well assessed.

The CVD process of monolayer graphene film synthesis involves fundamental steps including nucleation, growth and perfection, and the whole process is controlled by a number of factors such as the rates of hydrocarbon deposition and decomposition, surface diffusion of carbon sources, and the additive or etching reactions occurring at the growing edges.[14-15] Without knowing much detail at the atomic scale, the regime of growth can be characterized by the time evolution of single-crystal domain shapes, or the growth pattern, to infer the microscopic mechanisms. Once a stable graphene domain nucleates, the equilibrium shape of crystals can be determined from local-orientation-dependent edge energy densities following the conventional Wulff construction, in the condition that transport of atoms and relaxation of defective structures within the nuclei are sufficiently fast to reach thermal equilibrium.[16-17] However, as the incoming flux of precursors increases, the growth process deviates from the equilibrium, and edge-kinetics- or surface-diffusion-limited regimes are activated. At a low flux or deposition rate, the front of growth experiences the spatially-varying concentration field of active carbon sources that diffuse on the substrate. The growth process could thus become unstable if surface diffusion is insufficient to supply the additive reactants at domain edges, resulting in non-compact dendritic patterns.[16] While under high hydrocarbon flux and fast transport of carbon sources, reaction at the edges of growing crystalline domains becomes the rate-limiting factor, and the pattern becomes compact, well predicted by the kinetic Wulff construction from the growth rates. If the edge-kinetics-limited growth process is not much far from thermal equilibrium, one can assume the rate of growth to be proportional to the edge energy density.[18-19] This approach was applied to explain the experimentally observed shape evolution of single-crystal graphene domains during both growth and etching as its inverse process.[15]

By neglecting the atomic-level details of CVD growth, the evolutional process can be modelled by the phase-field method through a set of two coupled governing equations for the phase evolution and surface diffusion of carbon sources. The model has been successfully applied to study the effects of synthesis conditions on the development of single-crystal domain shapes,[13, 20] which was extended in this work to explore the growth and completion of monolayer graphene films from individual nuclei. In addition to the general features of a phase-field based crystal growth model, the anisotropic free energy functional is introduced for the front of growth as a functional of the local orientation.[21] The nucleation of single-crystal domains, however, cannot be studied using this model due to the absence of an energy barrier. This is not a critical issue for the current study as we are focusing on the growth and coalescence stages. The contrast between the surface diffusivities of carbon sources on the metal substrate and on/under the as-grown graphene domains can be considered in the model by introducing an order-parameter-dependent diffusivity. It should also be noted that the CVD process of graphene is a composite process where both growth and etching are active at edges of single-crystal domains, and the balance between them is controlled by the chemical driving force $\xi - \xi_{eq}$, where $\xi_{eq}$ is the equilibrium concentration of active carbon sources diffusing on the substrate. With these practical considerations, our model constructed with dimensionless parameters is validated by being able to reproduce a typical set of experimentally characterized growth patterns with both edge-kinetics-limited and surface-diffusion-limited features, which can be considered as a signature of the competitive growth and etching processes.

There are three characteristic time scales defined in the phase-field model for graphene growth – the time scales for (1) the deposition of a hydrocarbon precursor and its decomposition into reactive carbon sources, $1/F$, (2) the surface diffusion of the carbon sources to the growth front, $\sim l_D^2/D$ where $l_D$ is a characteristic length scale for the diffusive pathway, and (3) edge reactions, $\tau_\psi$. Competition between relevant processes determines the regime of CVD growth under specific synthesis conditions, which can be explored through the parameters $F$, $D$, and $\tau_\psi$ in the phase-field model. In recent experiments, the partial pressure of $CH_4/H_2$ and $Ar/H_2$ were tuned during the CVD process and a wide spectrum of grown patterns of single-crystal graphene domains was observed.[12] Our phase-field modeling results show that, at a low deposition and decomposition rate ($F$) from the carbon feedstock, the depletion zone with low precursor concentration ($\xi$) expands (**Figure 1**), indicating the lack of sufficient active carbon sources at the reactive edges, which further leads to the formation of the non-compact dendritic domains. From the spatial distribution of as-grown graphene (the order parameter field $\psi$) and the distribution of available carbon sources (the concentration field $\xi$), we find that it takes a long time to fill the deep and narrow gaps between neighboring domains in the dendrites, due to the short of carbon sources. Statistics show that the duration for the completion of a continuous graphene film, $\tau_f$, decreases dramatically with the deposition flux $F$, indicating a transition from the surface-diffusion-limited regime of graphene growth at a low flux to the edge-kinetics-limited regime at high fluxes (**Figure 1c**). The equilibrium concentration $\xi_{eq}$, which determines the chemical driving force for graphene growth, can be further tuned in the model to match the experimental conditions,[12, 14, 22] and a whole spectrum of domain shapes are uncovered where both growing and etching process are active. It should be noted that the incoming flux of active carbon sources, the chemical potential and edge reactivity could also be modified by the control of $H_2$ partial pressure, leading to changes in the parameters $F$, $\xi_{eq}$ and $\tau_\psi$ that could be further explored by using our model. In the experiments where the surface oxygen on the copper substrate is controlled during growth, significantly varying growth patterns were reported.[13] We conclude from our phase-field modeling results that by changing the controlling parameters $F$, $D$ and $\tau_\psi$, one can predict the evolution of domain shapes that align well with the experimental observation.

Although the evolution of single-crystal domains can be used to infer the microscopic mechanism of graphene growth, the total time for the completion of a continuous film that fully covers the substrate, $\tau_f$, is more practically concerned. First, this time scale defines the efficiency of material production. Secondly, the coalescence of individual single-crystal domains defines the microstructures of interfaces between them in the film, where defects could form due to the presence of lattice mismatch. The time scale to reach the full coverage thus determines the level of microstructural relaxation in the grain boundary region with lattice defects. From phase-field modeling one could probe these details by tracking the surface coverage of graphene on the substrate. From **Figure 2**, one can see that in the edge-kinetics-limited regime, the time evolution of individual graphene domains in the early-stage growth is independent of the domain size as the 'interaction' between neighboring domains resulted from the competition for carbon sources is absent. In contrast, in the surface-diffusion-limited regime, the growth of individual domains interferences with each other as the available carbon sources are limited, leading to significant reduction in the area of as-grown graphene. The growth rate of individual domains thus depends critically on the nucleation density or the distance between growing domains, which decreases as more nuclei are formed in the same area.

Based on the synthesis condition reported in Refs. [12-13], we summarize our phase-field modeling results in **Figure 3** to elucidate the effects of controlling parameters $F$, $D$, $\tau_\psi$ and $\xi_{eq}$. The results show that the growth rate increases with $F$, $D$ and the chemical driving force $\xi$-$\xi_{eq}$, but decreases with the reaction time scale $\tau_\psi$. The change of growth rate also signals a shift in the mode of growth. The rate for edge-kinetics-limited growth with compact graphene domains is significantly higher than that for the surface-diffusion-limited growth. However, there are common features shared between the two modes of growth, that is, the evolution of surface coverage $c(t)$ as a function of time $t$ first increases, following a quadratic profile, and then decreases gradually. Based on these modeling results, we consider the growth process as a three-stage process - (1) individual domain growth with no interference between neighboring domains, (2) competitive growth where neighboring domains approach each other, and (3) monolayer completion by filling the gaps between domains (**Figure 4**). The scaling behaviors of surface coverage $c(t) = t^r$ is distinct in these three stages (**Table 1**). As the deposition flux $F$ decreases from 1 to 0.01, we find that at the first stage, the exponent $r$ is almost a constant of ~2.0. This value of $r$ coincides with that for individual domain growth. The rate of carbon attachment to the as-grown domain scales linearly with the edge length $l_g$ and thus the surface coverage $c(t) \sim l_g^2$ increases with $t^2$, in the edge-kinetics-limited regime. For growth from multiple nuclei, the growth rate of $c(t)$ first increases and then decreases due to the competition between growing domains under the constraint of limited carbon sources. As a result, the value of $r$ decreases as growth proceeds, and becomes highly dependent on the value of $F$. At the third stage, the exponent is 0.68 for $F$ = 1, and 0.19 for $F$ = 0.01. These results suggest that at the early stage of the growth rate is less sensitive to the mode of growth, which can be explained by the fact that although the edge-kinetics-limited and surface-diffusion-limited growth modes yield quite different domain shapes as demonstrated by the distribution of order parameter $\psi$, the concentration field is less sensitive to $F$, which conforms to the coarsened shape of graphene domains (**Figure 1a**). To measure the irregularity of the concentration field at the growing fronts of graphene domains, we consider the edge profiles of grown domains as fractal objects. Regularization dimension (RD) analysis using the FracLab toolbox shows that for typical compact and dendrite domains (with parameters $F$ = 1, 0.005), the RD values for the graphene domain (the $\psi$ field) ranges from 2.5 to 2.0, but the discontinuity in the profiles of the carbon source concentration (the $\xi$ field) is almost independent on the value of $F$, with RD = ~2.0-2.1. Consequently, for the dendritic mode of growth, the active sites for edge reaction are more relevant to the coarsened edge profile that is quantified by the discontinuity in the concentration field instead of that in the order parameter field, and thus the growth rate at the early stage is mode-insensitive. On the other hand, the completion of graphene monolayer, at the last stage of growth, is highly sensitive to the mode of growth. The rate of completion for the continuous graphene films is much higher for the compact domains (the edge-kinetics-limited regime) than that for the dendrites (the surface-diffusion-limited regime).

To validate these theoretical arguments, we track the evolution of domain shapes and surface coverage in the CVD process. The experimental setup follows our previous work.[12, 23] The data clearly demonstrates the three stages of graphene growth in both the edge-kinetics-limited and surface-diffusion-limited modes, which aligns with our predictions from phase-field modeling (**Figure. 5**). The evolution of surface coverage increases first rapidly and then decays in the final stage, and the overall growth rate of compact domains is much higher than that for the dendritic ones. The misalignment of graphene domains in the final stage of growth as observed in the experiments may be attributed to the irregularity in the nucleation process and the reduced surface diffusivity of large-area domains on the substrate.

Understandings of the evolutional dynamics CVD growth could guide the practical optimization of synthesis conditions for fabricating large-size single-crystal graphene monolayers, for which usually two strategies could be taken. One could grow a large single-crystal domain from a single nucleus, or a continuum monolayer through the coalescence of an ensemble of individual domains nucleated at multiple sites. We now compare the time required to grow graphene films with the same area by these two procedures. Moreover, the quality of as-grown graphene monolayers is defined by the final process of coalescence where the neighboring domains merge. For compact domains, seamless, defect-free interfaces could be produced if the domains are well aligned to the same orientation, or local movement and deformation are allowed to relax the potential mismatch between the domains.[5, 24-25] However, for the film formed from fractal patterns, our modeling results show that in the final stage, there are a significant number of pores within the films, which may be difficult to be completely healed by the deposited hydrocarbon sources or structural annealing. As a result, local defects could remain the continuous film. The results suggest that strategies to achieve efficient graphene growth with high rates in the two regimes are different. In the edge-kinetics-limited regime, the growth rate is sensitive to the density of nucleation sites. As a result, reducing the density of nucleation enhances the yield of compact graphene domains, and growing a large-size, single-crystalline domain from a single nucleus is recommended. While in the surface-diffusion-limited condition, the duration to complete the film growth is insensitive to the density of nucleation, and thus if we are growing dendritic domains, the cost to keep a low nucleation density should be taken into account for the limited rewards. In addition, it should be noted that the formation of continuous graphene monolayer does not mean the full surface coverage ($\tau = \tau_f$). To demonstrate this fact, we calculate the time for the individual domains to percolate by coalescence, $\tau_p$, and the ratio $\tau = \tau_p/\tau_f$, which show that the ratio varies between 0.4 and 0.8, depending on the growth mode that is controlled by the synthesis conditions (**Table 2**). The results clearly indicate the difficulty for the active carbon sources to diffuse on the substrate into the narrow gaps between neighboring domains and become attached to the reactive edges at the final stage of growth ($\tau_f < \tau < \tau_f$), especially when the flux $F$ is low. In practice, the starvation of active carbon sources could be relaxed through the back-side diffusion across a thin substrate as demonstrated in a recent study.[26]

In brief, we modelled the CVD growth process of monolayer graphene through phase-field modeling. The results unveil a clear correlation between the synthesis conditions (the deposition and decomposition rate of precursors, the surface diffusivity of carbon sources, the chemical driven force, the edge reactivity) and the rate of growth, which can be attributed to the shift in the regime of graphene growth from edge-kinetics-limited to surface-diffusion-limited, as identified from the growth patterns. The conclusion may also apply for other 2D materials such as the hexagonal boron nitride, where a similar time evolution of film coverage was identified.[27]

## Supporting Information

The Supporting Information is available free of charge on the ACS Publications website at DOI: 10.1021/acs.XX.XX.

Computational and experimental details; validation of phase-field modeling results by comparing with the experimentally characterized pattern evolution.

## Acknowledgments

This work received no financial support. The computation was performed on the Explorer 100 cluster system of Tsinghua National Laboratory for Information Science and Technology.

# Figures, Tables and Captions

**Table 1** The scaling factor *r* that characterizes the evolution of graphene coverage on the substrate $c(t) = t^r$ in the three characteristic stages as identified from data shown in **Figure 4**.

|       | F = 1  | F = 0.5 | F = 0.1 | F = 0.05 | F = 0.01 |
|-------|--------|---------|---------|----------|----------|
| $r_1$ | 1.9895 | 1.9873  | 2.0718  | 2.1078   | 2.0285   |
| $r_2$ | 1.9683 | 1.9050  | 1.9328  | 1.8686   | 1.5910   |
| $r_3$ | 0.6763 | 0.5871  | 0.4355  | 0.3120   | 0.1864   |

**Table 2** The ratio $\tau = \tau_p/\tau_f$ defined between the time required to form a percolated film from individual domains, $\tau_p$, and that for the completion of full coverage, $\tau_f$, obtained from phase-field modeling results with specified parameters $F$, $\tau_\psi$ and $\xi_{eq}$.

| $\tau_\psi$ \ $F$ | 1 | 0.5 | 0.05 | 0.005 |
|---|---|---|---|---|
| 1 | 0.7888 | 0.6984 | 0.5777 | 0.3968 |
| 0.5 | 0.7485 | 0.6812 | 0.5220 | 0.4176 |
| $\xi_{eq}$ \ $F$ | 1 | 0.1 | 0.01 | 0.001 |
| 1 | 0.8154 | 0.8126 | 0.8122 | 0.8122 |
| 0.01 | 0.7881 | 0.6953 | 0.6393 | 0.4804 |

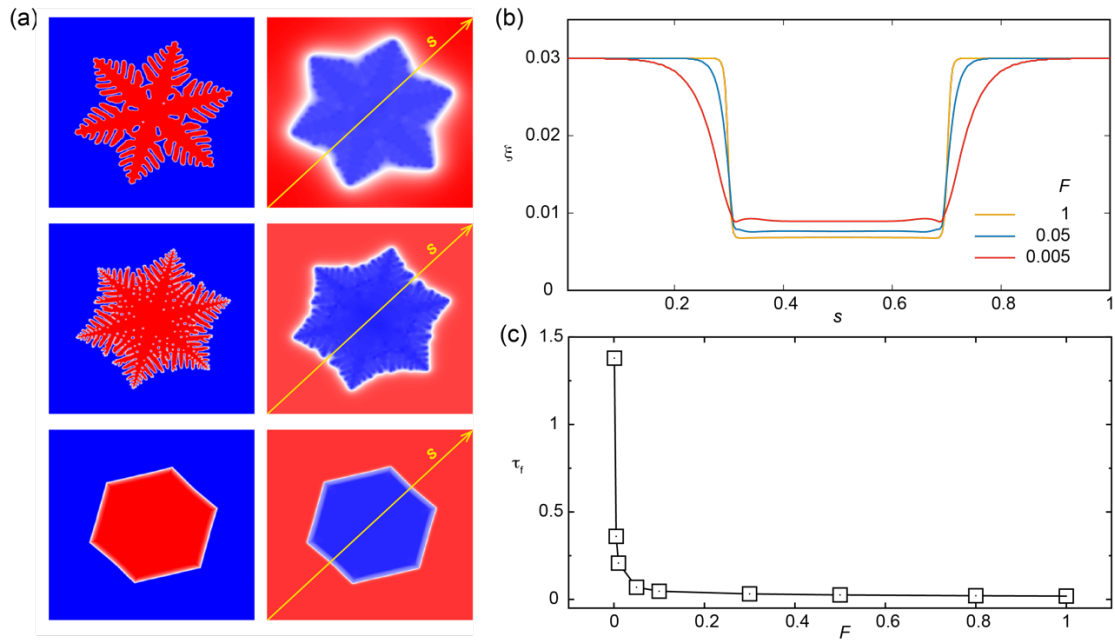

**Figure 1** (a) The order parameter (left) and the concentration of active carbon sources on the substrate (right). The distribution of concentration along the yellow path with coordinate $s$ is plotted at different flux $F$ in panel (b). (c) The time required for the graphene film to fully cover the substrate, $\tau_f$, plotted as a function of $F$, modelled with parameters $D = 60$, $\xi_{eq} = 0.03$, $\tau_\psi = 2$.

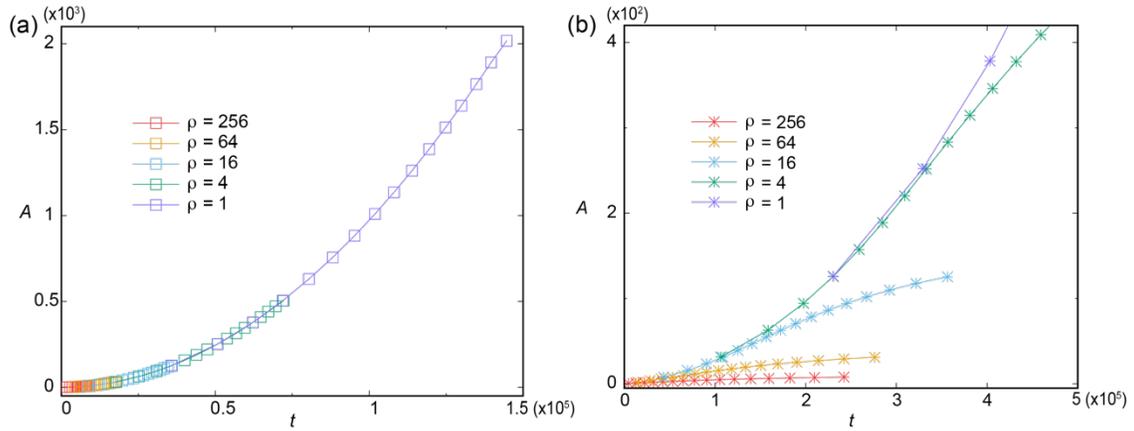

**Figure 2** The evolution of domain size *A* measured at different areal densities $\rho$ of nucleation, which demonstrates the cooperative effect between neighboring graphene domains growing in the edge-kinetics-limited (panel a, *F* = 1) and surface-diffusion-limited regimes (panel b, *F* = 0.01), respectively (*D* = 60, $\xi_{eq}$ = 0.03, $\tau_\psi$ = 2). The size of graphene domains *A* is measured in the characteristic length scale $l_c$, where $l_c^2 = D\tau_\psi$, and the density $\rho$ is measured by the number of nucleated domains in the model.

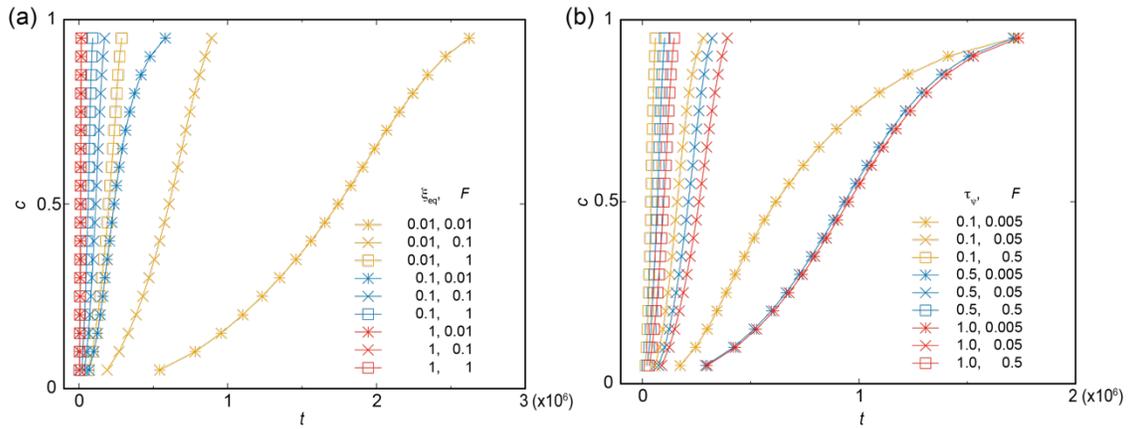

**Figure 3** The evolution of surface coverage modelled according to the experimental conditions, where (a) the partial pressure of $CH_4/H_2$ and $Ar/H_2$ are tuned, resulting in changes in $\xi_{eq}$, $F$, and $\tau_\psi$ (in this plot we fix $\tau_\psi = 2$), and (b) surface oxygen is controlled to modulate $\tau_\psi$, $F$, and $D$ (in this plot we fix $D = 60$).

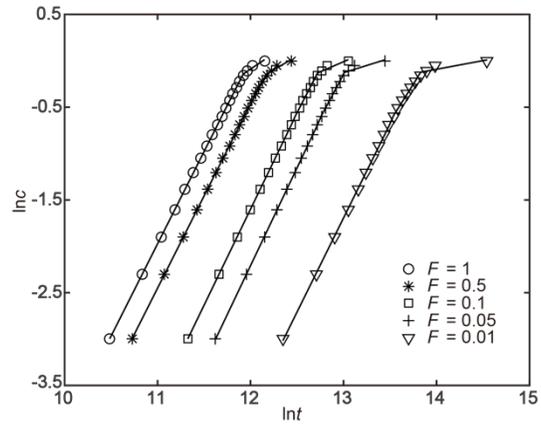

**Figure 4** Scaling analysis for the time evolution of graphene surface coverage with parameters $F$ = 0.01-1, $D$ = 60, $\xi_{eq}$ = 0.03 and $\tau_\psi$ = 2.

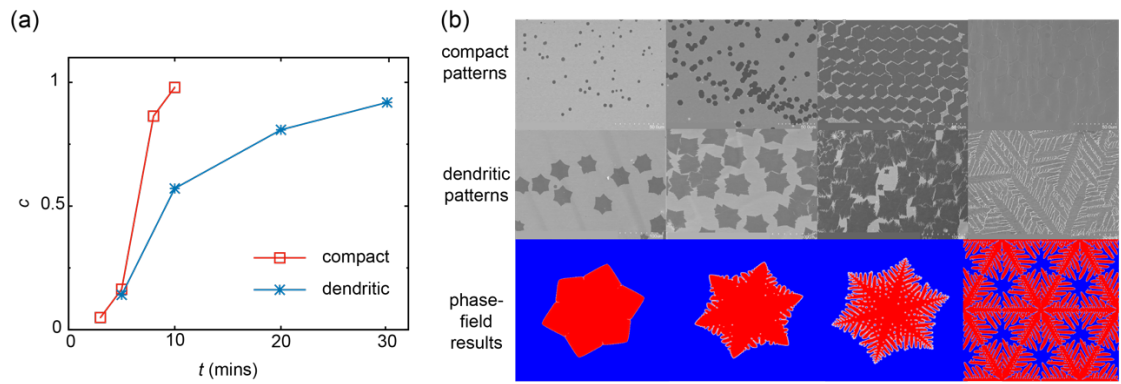

**Figure 5** (a) The evolution of graphene surface coverage measured in experiments (b) The growth patterns obtained from experiments and phase-field modeling.